\documentclass[sigconf,nonacm]{acmart}
\usepackage{tcolorbox}
\usepackage{graphicx}
\usepackage{enumitem}
\usepackage{float}
\usepackage[ruled,vlined]{algorithm2e}
\usepackage{placeins}
\usepackage{tabularx}
\usepackage{algpseudocode}
\usepackage{pifont}
\usepackage{bbding}
\usepackage{balance}
\usepackage{multirow}

\AtBeginDocument{
  }

\renewcommand\footnotetextcopyrightpermission[1]{}
\begin{document}

\title{When and How to Ask: Dynamic Preference Elicitation Strategies for Conversational Recommendation}

\author{Feng Xia}
\affiliation{%
  \institution{University of Sheffield}
  \city{Sheffield}
  \country{United Kingdom}
}
\email{fxia8@sheffield.ac.uk}

\author{Shuo Zhang}
\affiliation{%
  \institution{Bloomberg}
  \city{London}
  \country{United Kingdom}}
\email{szhang611@bloomberg.net}

\author{Xi Wang}
\affiliation{%
  \institution{University of Sheffield}
  \city{Sheffield}
  \country{United Kingdom}
}
\email{xi.wang@sheffield.ac.uk}

\renewcommand{\shortauthors}{Feng Xia, Shuo Zhang, \& Xi Wang}

\begin{abstract}
 Conversational Recommender Systems (CRSs) are interactive systems that use multi-turn natural language dialogue to understand evolving user preferences and provide personalized recommendations. To achieve this goal, CRSs rely on preference elicitation strategies to actively gather informative preference cues from users; however, the timing and selection of these strategies during a conversation remain largely unexplored. While many existing studies emphasize eliciting explicit item attributes and tend to adopt relatively static elicitation strategies, the use of item-based preference elicitation and how it varies across different dialogue stages remains less explored. In this work, we conduct a systematic investigation of preference elicitation strategies from a stage-aware perspective. We provide empirical evidence that optimal preference elicitation strategies are stage-dependent and context-sensitive: attribute-based inquiries are effective in early stages, while item-based strategies become superior as preferences refine. To support this paradigm, we introduce \textbf{InPE}, a dataset enriched with fine-grained annotations for elicitation necessity and strategy selection.  
 With this dataset, we propose \textbf{COPE} (\textbf{CO}nversational \textbf{P}reference \textbf{E}licitation via Mixture of Experts), a novel architecture for strategy modeling. Extensive offline evaluation on our dataset indicates that context-aware preference elicitation strategies are beneficial for conversational recommendation. In addition, the analysis of the predicted strategies uncovers consistent stage-wise tendencies in dialogue progression, providing empirical evidence of common interaction patterns in conversational recommendation systems. Our dataset is available at \url{https://github.com/juanfacabian/InPE}.
\end{abstract}



\maketitle
\section{Introduction}
Conversational Recommender Systems (CRSs) deliver personalized recommendations through natural language interactions with users. A core advantage of CRSs lies in their ability to actively collect user preferences during a conversation, enabling more accurate and adaptive personalized recommendations. However, capturing users' evolving preferences and determining when and how to elicit them in a conversation remain fundamental challenges in building effective CRSs.

Traditional recommender systems infer user preferences primarily from historical user behavior \cite{rendle2012bpr}, sometimes augmented with user-centric knowledge graphs \cite{wang2019kgat}. While effective in many scenarios, these approaches are significantly limited by noisy and sparse collaborative signals. For example, users may click on items they do not actually like. In contrast, CRSs exhibit greater proactiveness and flexibility in collecting user preferences rather than relying solely on human feedback.
One fundamental way to collect user preferences in CRSs is through preference elicitation methods. Unlike approaches that utilize models to infer user preferences (e.g., \cite{deng2021unified}), CRSs can proactively elicit preferences by adapting their dialogue policies, i.e., strategies that determine which actions to take based on the current dialogue state. For example, when users have not specified any movie interests, a CRS may explore their preferences by asking questions such as \emph{``What kind of movies do you like?''}

\begin{figure}[t]
  \centering
  \includegraphics[width=1.0\linewidth]{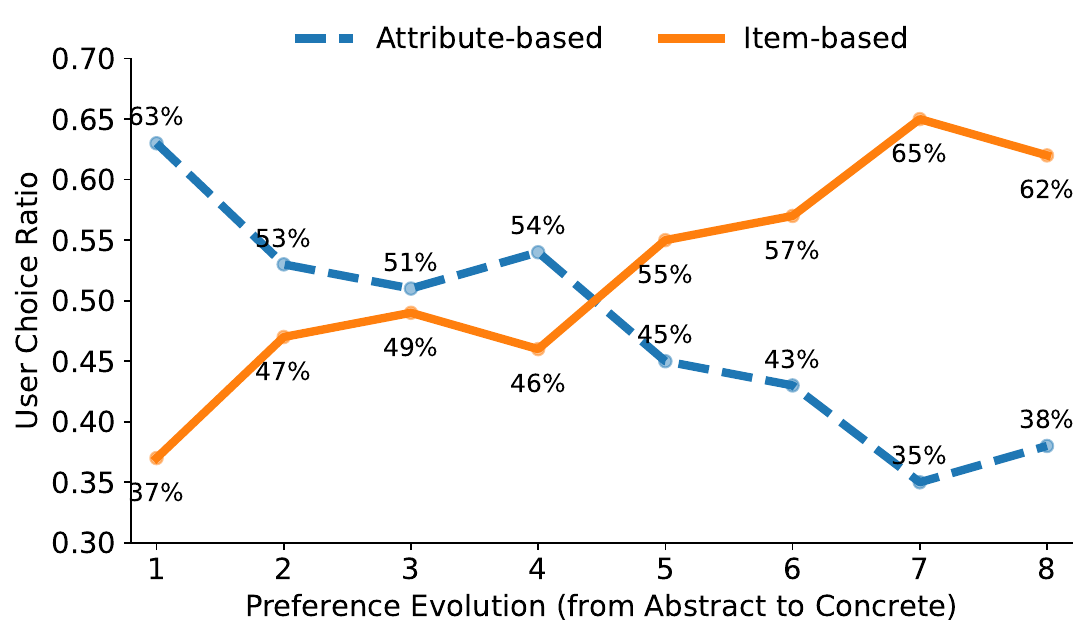}
  \caption{User selection ratios of preference elicitation strategies across dialogue stages. The x-axis shows the progression from abstract to concrete preferences (Stages~1--8), and the y-axis denotes the proportion of attribute-based and item-based strategy selections.}
  \label{fig:stage}
\end{figure}

According to the classification in \cite{gao2021advances}, existing preference elicitation methods can be divided into two categories: attribute-based methods \cite{priyogi2019preference} and item-based methods \cite{mccarthy2010experience,mccarthy2004dynamic}. Attribute-based methods elicit preferences by asking attribute questions, such as preferred movie genres \cite{zou2020towards}. Accordingly, much prior work has focused on how to generate and select these questions. For instance, ~\cite{priyogi2019preference} and ~\cite{zou2020towards} focus on the selection and ranking of elicitation questions from predefined pools, whereas ~\cite{lei2020estimation} and ~\cite{ren2021learning} emphasize attribute importance modeling, prioritizing item attributes based on users’ historical interactions. Another approach to preference elicitation is the use of item-based methods, which present concrete items for users to compare. Many studies employ item-based methods for preference elicitation, e.g., ~\cite{bledaite2015pairwise}, which asks users to compare two items, and ~\cite{yu2019visual} provides multiple items for users to choose from.

Despite the effectiveness of these methods, most CRSs rely on a single elicitation strategy and apply it uniformly throughout the conversation. In practice, attribute-based elicitation is often treated as the default, while item-based strategies are used in limited or ad hoc scenarios. However, user preferences naturally evolve as the dialogue progresses, from early exploratory stages with abstract requirements to later stages with more concrete and refined preferences. As illustrated in Figure~\ref{fig:stage}, users exhibit different selection tendencies toward elicitation strategies across dialogue stages, suggesting that no single strategy is universally optimal.

Most prior work has focused on deciding whether a system should ask questions or make recommendations, while paying limited attention to how different elicitation strategies perform across dialogue stages. As a result, the stage-dependent applicability of preference elicitation strategies remains insufficiently understood, and current CRSs lack explicit mechanisms for dynamically selecting appropriate strategies during a conversation.

To systematically investigate the role of preference elicitation strategies in CRSs and address these gaps, we formulate the following research questions:
(1) \textbf{RQ1:} Is a single commonly used preference elicitation strategy sufficient to elicit user preferences across diverse dialogue stages in CRSs?
(2) \textbf{RQ2:} Do preference elicitation strategies exhibit stage-dependent applicability across different dialogue stages?
(3) \textbf{RQ3:} Does explicitly modeling elicitation strategies lead to improved CRS performance?

To address RQ1 and RQ2, we conduct a user study to examine the effectiveness of different elicitation strategies associated with comprehensive dataset annotations across dialogue stages. To investigate RQ3, we analyze the relationship between subjective user feedback and elicitation strategies, and propose a novel \textbf{CO}nversational recommendation system architecture for \textbf{P}reference \textbf{E}licitation via Mixture of Experts (\textbf{COPE}), which explicitly models heterogeneous preference elicitation strategies within a unified framework. 

Our main contributions are as follows:
(1) We empirically validate that the effectiveness of elicitation strategies is dynamic. We show that shifting from attribute-based to item-based strategies aligns with the natural transition from abstract to concrete user preferences.
(2) We introduce \textbf{\textit{InPE}}, a dataset that transforms the INSPIRED corpus by adding fine-grained annotations across the elicitation necessity and preferred strategy choice, and user preferences of responses. This provides the first benchmark for training elicitation strategy-aware CRS systems.
(3) We develop \textbf{COPE}, a strategy-aware CRS. By treating conversation as a sequential decision-making process over a structured action space, COPE utilizes a Mixture-of-Experts (MoE) approach to master the timing and form of preference elicitation, consistently outperforming strong LLM-based baselines.

\section{Related Work}
In this section, we explore related work in two areas. We review \textit{preference elicitation approaches} and then discuss \textit{proactiveness in conversational recommendation}, highlighting the shift from passive history tracking to system-initiated interactions to improve user experience.

\subsection{Preference Elicitation in Conversational Recommender Systems}
Following the taxonomy introduced in \cite{gao2021advances}, we categorise existing user preference elicitation methods into two primary streams, namely \textit{attribute-based methods} \cite{priyogi2019preference} and \textit{item-based methods} \cite{mccarthy2010experience, mccarthy2004dynamic}. Attribute-based methods elicit user preferences by asking attribute-related questions, such as preferred movie genres or book authors \cite{zou2020towards} or asking users to critique or justify their preferences on specific item attributes \cite{lei2020estimation}. For example, \citet{priyogi2019preference} and \citet{zou2020towards} elicited user preferences by selectively asking questions from a question pool that is constructed using entities in the item descriptions and reviews while optimizing the question selection strategies. Similarly, \citet{lei2020estimation} and \citet{ren2021learning} prioritized item attributes by relevance, selecting the most informative ones to query users.

Conversely, item-based methods offer a group of complementary strategies. By presenting items for evaluation, comparison, or selection, these methods provide a straightforward way for users to disclose preferences that are often difficult to articulate in natural language. In addition, asking users to select items among several options yields contrastive feedback, which has proven effective for user preference modeling\cite{yu2019visual,hu2022learning}. For example, \citet{bledaite2015pairwise} derived pairwise preferences by asking users to choose between two items rather than rating them individually. Moreover, presenting complete items allows a CRS to capture holistic preferences. However, despite the potential benefits, most existing systems rely on a single, fixed strategy \cite{xie2021comparison}. The potential for dynamically selecting elicitation strategies remains underexplored, limiting the system's ability to adapt across different stages of the dialogue. Addressing this limitation necessitates a shift towards proactive conversational recommender systems. Unlike static approaches, proactive systems are designed to dynamically determine the optimal preference elicitation strategy, which can switch between attribute queries and item presentations, so as to actively guide the conversation toward a successful recommendation. 

\subsection{Proactiveness in Conversational Recommendation}
Traditional conversational recommendation systems mainly rely on dialogue history to infer user preferences. However, passive context tracking is often insufficient for capturing complex user intent. To improve interaction efficiency, growing research contributions have shifted toward proactive conversational recommender systems, where the system takes the initiative to elicit user preferences rather than relying solely on users' input. The proactiveness of a CRS is reflected through various system intervention strategies, such as \textit{attribute querying} to narrow the search space \cite{ren2021learning}, \textit{exploratory recommendations} to gauge user interest \cite{christakopoulou2016towards}, \textit{proactive explanations} to verify user preference alignment \cite{hernandez2023explaining}, and \textit{active item usage exploration} to clarify their needs\cite{kostric2024generating}. 

With the system constantly deciding \textit{which} intervention to apply or whether to make a final recommendation, based on the evolving state of the conversation, this task often involves a sequence of decision-making. Accordingly, prior works have often adopted Reinforcement Learning (RL) frameworks. For example, \cite{deng2021unified} proposed a graph-based RL formulation with dynamically weighted structures to track evolving user preferences, while MCMIPL \cite{zhang2022multiple} improves elicitation efficiency by proactively presenting multi-choice questions to capture user complex interests. However, despite these advances, a key limitation remains: existing proactive policies lack fine-grained, context-aware strategy switching, often relying on fixed-stage heuristics or single-type proactive behaviors \cite{sun2018conversational}. To address this gap, this study develops resources grounded in real human preferences over preference elicitation strategies across diverse dialogue contexts, enabling the learning and optimization of fine-grained, context-aware proactive strategy switching in conversational recommendation systems.

\section{Preliminary User Study}
To investigate the trade-off between using single attribute-based or item-based methods versus multiple preference elicitation strategies (RQ1) and to explore the distinct value of these strategies across different conversational stages (RQ2), we conducted a preliminary user study. We recruited 328 participants through a crowdsourcing platform\footnote{\url{https://app.cloudresearch.com/}} to complete a survey hosted online\footnote{\url{https://www.qualtrics.com/}}. To ensure eligibility, participants first completed a screening questionnaire regarding their demographic information and prior experience with AI systems. Then, they engaged in several predefined conversational scenarios, where they were asked to select their preferred system responses. As each potential system response reflects a specific preference elicitation strategy, the participants' selections provided empirical evidence of user preferences for different strategies in varying contexts.

\subsection{Preliminary Observations}
\subsubsection{RQ1: Sufficiency of Single Elicitation Strategies}
To determine whether a single preference elicitation strategy is sufficient across varying dialogue stages, participants evaluated the effectiveness of responses across multiple dialogue scenarios, where each response was generated using a different strategy (attribute-based or item-based methods). Results indicate that a static approach is rarely perceived as sufficient. Regarding attribute-based elicitation, 54\% of participants reported it was effective in most situations, whereas only 21\% found it sufficient across all situations, and 25\% considered it effective in only a few cases. Item-based elicitation yielded similar results, with corresponding proportions of 59\%, 25\%, and 16\%, respectively.  These findings confirm that no single elicitation strategy can robustly support user preference expression across diverse dialogue contexts, highlighting the need for a hybrid or dynamic approach.

\subsubsection{RQ2: Stage-Dependent Patterns in Preference Elicitation}
Next, we investigated user preferences for elicitation strategies across distinct conversational stages. We designed eight dialogue scenarios reflecting three main dialogue stages (preference exploration, preference refinement, and post-recommendation preference readjustment). These scenarios spanned user needs ranging from initially vague to clearly defined. Figure~\ref{fig:stage} shows the proportion of user choices between attribute-based and item-based elicitation strategies at each preference stage. In the early stages, attribute-based elicitation dominates user choices. However, as the dialogue progresses and preferences become more concrete, the use of item-based elicitation increases steadily and eventually surpasses attribute-based strategies in the later stages. This shift reveals a stage-dependent pattern: attribute-based strategies are more effective for early-stage preference exploration, whereas item-based strategies become more suitable once users have formed clearer and more specific preferences. 

Collectively, these findings suggest that the optimal preference elicitation strategy is not static but varies across dialogue stages. However, most existing CRS datasets do not provide annotations for elicitation strategies at this granularity level. This limitation motivates our subsequent effort to annotate data to support stage-aware analysis and modeling.

\section{Data Annotation and Augmentation: The InPE dataset}
\label{sec:data_annotation}
To enable a fine-grained analysis of preference elicitation strategies across different dialogue stages, we introduce \textbf{InPE} (\textbf{In}SPIRED \textbf{P}reference \textbf{E}licitation), an annotated and augmented version of the INSPIRED~\cite{hayati2020inspired} dataset. While the original INSPIRED dataset has been widely used \cite{he2023large,wang2022towards,guo2023towards}, providing a robust foundation, which contains 999 multi-turn dialogues with an average of 21.15 turns per dialogue, it lacks strategy-level annotations. We address this by enriching the data with explicit strategy labels obtained via crowd-sourced annotation\footnote{\url{https://labelbox.com/}}.

\subsection{Annotation Schema and Procedure}
\label{subsec:annotation}

\begin{figure*}[t]
  \centering
  \includegraphics[width=0.80\linewidth,
    trim=0.5cm 20cm 0.5cm 0.5cm,
    clip
  ]{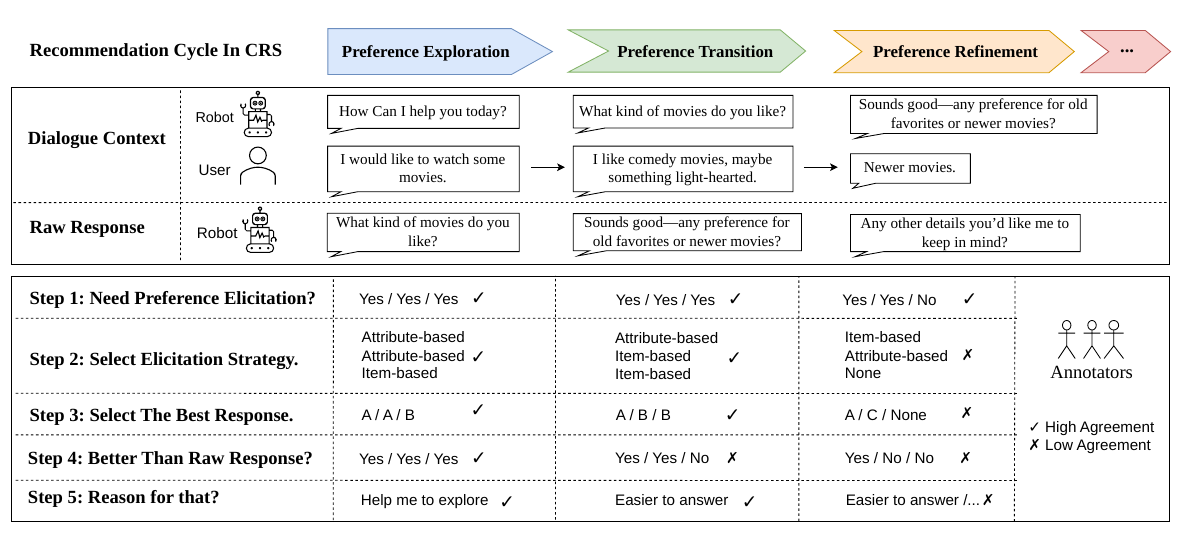}
  \caption{Overview of the step-by-step annotation workflow.}
  \label{fig:flow}
\end{figure*}

To ensure both efficiency and quality, we adopted a semi-automated annotation framework combining Large Language Model (LLM) filtering with human verification. We first employed an LLM (i.e., Qwen3-32B \cite{qwen3technicalreport}) to scan the dataset to identify dialogue turns with potential preference elicitation needs. We then devised a turn-level annotation schema for human annotators to \textit{verify} and \textit{label} these candidate turns. To ensure high-quality data, annotators were required to review the instructions and pass a qualification quiz as a prerequisite for participation. Figure~\ref{fig:flow} presents an overview of the sequential workflow that the annotations follow during the task, from the preference elicitation need identification to the choice of preferred elicitation strategies and responses. Annotators also comparatively evaluate the response generated using the selected strategy and decide if it outperforms the original response, with associated explanations.

For each dialogue turn identified by the LLM filter, annotators first review the context to validate whether preference elicitation is genuinely required. Turns lacking this intent are excluded from further annotation. For eligible turns, annotators classify the elicitation strategy used by the system as \emph{attribute-based}, \emph{item-based}, or \emph{hybrid}. Note that hybrid strategies indicate that the system considers both items and attributes for joint preference elicitation. Next, for each eligible turn, the system provides multiple candidate responses generated under the selected strategy, and annotators choose the most appropriate response.
Annotators then assess whether the selected response constitutes an improvement over the original system utterance. If no improvement is identified, annotation for that turn ends; otherwise, annotators record the corresponding reasons for the improvement and submit the annotation. To ensure reliability, as illustrated in Figure~\ref{fig:flow}, each dialogue turn is annotated independently by three annotators, with the final label determined via majority voting.
\begin{table}[t]
\centering
\caption{Overall annotation statistics for preference elicitation and strategy selection.}
\label{tab:annotation_stats}
\small
\begin{tabular}{lccc}
\toprule
\textbf{Question} & \textbf{Label} & \textbf{Count} & \textbf{Proportion (\%)} \\
\midrule
Q1: Preference elicitation & Yes & 2063 & 60.02 \\
                           & No  & 1374 & 39.97 \\
\midrule
Q2: Strategy type          & Attribute & 497 & 24.09 \\
                           & Item      & 595 & 28.84 \\
                           & Hybrid    & 969 & 46.97 \\
\bottomrule
\end{tabular}
\end{table}

\subsection{Data Annotation Analysis}
\paragraph{Annotation Statistics.}
As outlined in the procedure, the LLM served as the initial filter to screen the INSPIRED dataset. Of the 10,576 total dialogue turns, 3,437 turns (32.5\%) were flagged by the LLM as candidates for preference elicitation. Subsequent human validation confirmed that 60.07\% of such flagged turns require preference elicitation. Table~\ref{tab:annotation_stats} summarizes the statistics for manual annotations of preference elicitation and strategy selection.
Among these, hybrid strategies, which perform preference elicitation by simultaneously considering items and attributes, account for the largest proportion (47.03\%), followed by item-based (28.85\%) and attribute-based (24.12\%) strategies. This distribution suggests that effective preference elicitation often involves the dynamic combination of multiple preference elicitation cues rather than a single, isolated strategy.

\paragraph{Case Study} To further illustrate InPE, we sample an example from its test set that illustrates how the system adjusts its strategy selection across conversation turns. Table~\ref{tab:case_dialogue} presents such an example. As the dialogue progresses, the system can choose different actions, shifting from \emph{Preference Elicitation}, then to \emph{Recommendation}, and to \emph{General Interaction} in later turns. In particular, in the Preference Elicitation stage, the system is annotated to employ an \emph{attribute filling} strategy to identify broad user preferences, such as general movie genres. With the initial preferences, the system changes its strategy to \emph{item select}, offering specific items like \emph{Jojo Rabbit} to refine the user's nuanced taste. Ultimately, the system successfully completes the Recommendation task by suggesting \emph{Hunt for the Wilderpeople} and then starts with general interaction or light chit-chat. Overall, our InPE dataset provides comprehensive annotations to enable the development of advanced CRS systems over fine-grained proactiveness cues.

\begin{table}[t]
\centering
\caption{Case Study: Example Dialogue from the Test Set Illustrating Stage-aware Strategy Use}
\label{tab:case_dialogue}
\renewcommand{\arraystretch}{1.0}
\small
\begin{tabular}{p{0.3\linewidth} p{0.6\linewidth}}
\toprule
\textbf{Dialogue Stage} & \textbf{Dialogue Trace} \\
\midrule
\textbf{Preference Elicitation} &
\textbf{dialogue context:} User requests a movie recommendation at the start of the conversation.\\
& \textbf{Strategy:} attribute\_filling \\
& \textbf{Assistant:} What kinds of movies do you usually enjoy? \\
& \rule{1.0\linewidth}{0.3pt} \\
& \textbf{dialogue context:} User wants something lighthearted; mentions \emph{What We Do in the Shadows} \\
& \textbf{Strategy:} item\_select\\
& \textbf{Assistant:} Among these options—\emph{What We Do in the Shadows} and \emph{Jojo Rabbit}—which one would you prefer? \\
\midrule
\textbf{Recommendation} &
\textbf{dialogue context:} User expresses specific preference for quirky fantasy; likes director \emph{Taika Waititi} \\
& \textbf{Assistant:} I recommend \emph{Hunt for the Wilderpeople}. \\
\midrule
\textbf{General Interaction} &
\textbf{User context:} ... \\
\bottomrule
\end{tabular}
\end{table}

\begin{figure}[t]
  \centering
  \includegraphics[width=.8\linewidth]{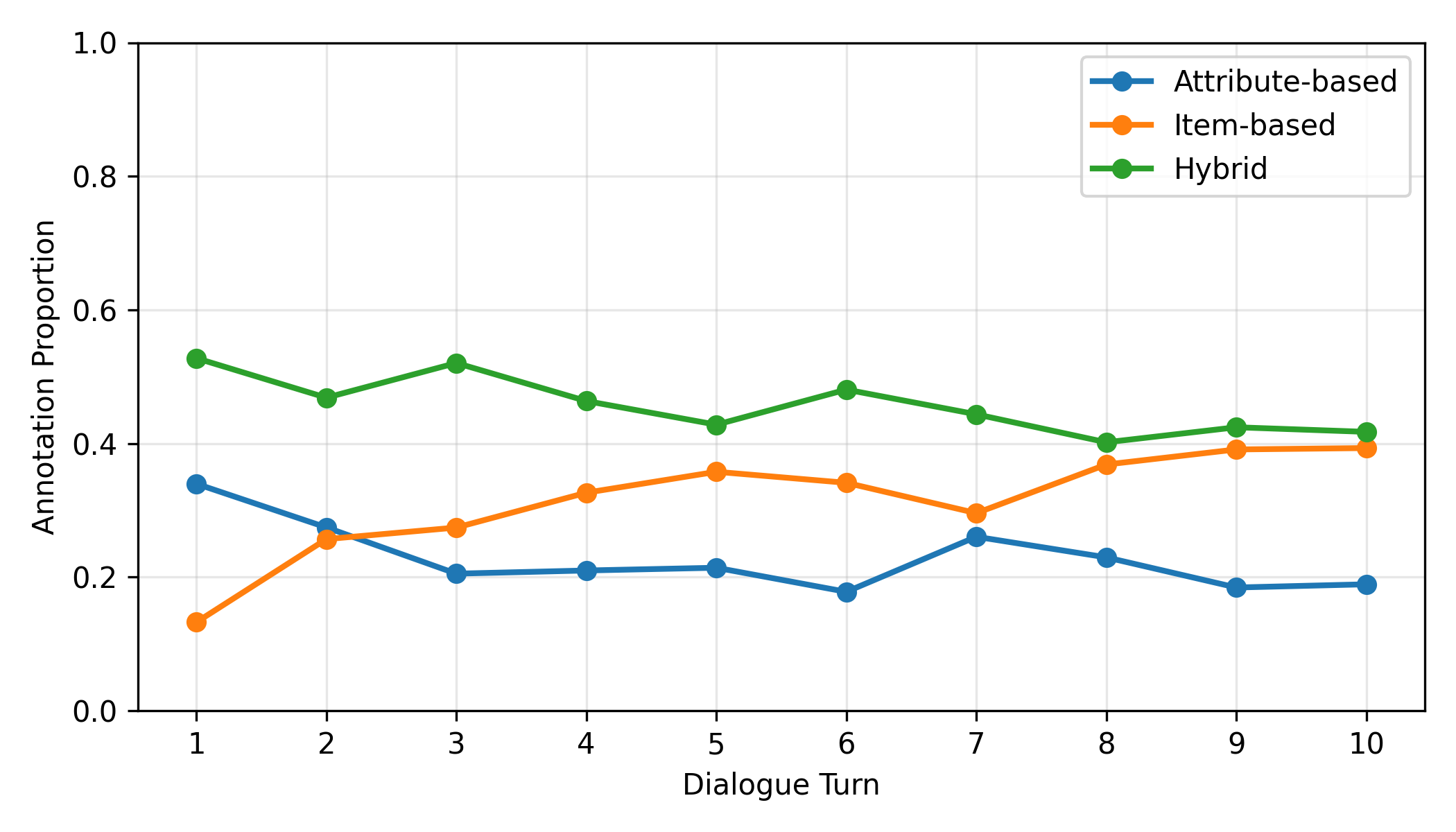}
  \caption{Proportion of different elicitation strategies in annotated dialogue turns.}
  \label{fig:stage_strategy}
\end{figure}

While Table~\ref{tab:annotation_stats} provides an overview of strategy usage and Table~\ref{tab:case_dialogue} further illustrates with the example, it does not show how elicitation strategies evolve over the stages of a dialogue. To analyze these dynamics, Figure~\ref{fig:stage_strategy} illustrates the normalized distribution of elicitation strategies at each turn. The results reveal distinct stage-dependent patterns that strongly align with the findings from our preliminary user study. Specifically, attribute-based elicitation is more prominent in the early turns, whereas the prevalence of item-based elicitation steadily increases as the dialogue progresses and user preferences become more concrete. Meanwhile, hybrid strategies maintain a consistent presence throughout the interaction, underscoring their versatile utility across all conversational stages.

Beyond analyzing strategy distributions, we further investigated whether employing strategy-aware preference elicitation methods can lead to improved user experience. Table~\ref{tab:strategy_better} shows the results of human preference evaluation comparing strategy-aware preference elicitation responses against the original system responses. For each turn, participants indicated whether the strategy-aware variant was superior or inferior to the original one. The results demonstrate a consistent and substantial preference for strategy-aware methods across the three categories; participants favored the strategy-aware responses in the vast majority of cases, with preference rates ranging from 89.34\% to 95.5\%.

Given that these results rely on annotated strategy and elicitation labels, we further assess the reliability of the annotation results. We quantify inter-annotator agreement using Krippendorff’s $\alpha$ \cite{krippendorff2018content}, a standard metric for data annotation tasks. Table~\ref{tab:consistency} presents the agreement scores: we observe an $\alpha$ of 0.244 for the preference elicitation decision (Q1) and 0.531 for the strategy type classification (Q2). These values indicate distinct levels of subjectivity inherent in each task. The moderate agreement for strategy classification (0.531) suggests that once an elicitation intent is identified, annotators can distinguish between strategy types (e.g., attribute vs. item) with reasonable consistency. 
Conversely, the lower agreement for the initial decision (0.244) highlights the inherent subjectivity and ambiguity in determining whether a conversational turn implicitly requires preference elicitation. To mitigate this variance, disagreements were resolved via majority voting. In cases where no majority was reached, an expert annotator reviewed the sample to assign the final label. This was particularly crucial in the transitional stage, where user preferences shift from exploration to refinement, as multiple strategies can serve complementary roles in this ambiguous context.

Overall, the above analyses indicate that preference elicitation in conversational recommendation is dynamic and stage-dependent. Different elicitation strategies exhibit distinct usage patterns across dialogue stages, and incorporating strategy awareness consistently improves user-perceived response quality. 

These observations suggest that an effective proactive CRS must do more than simply generate text; it must actively navigate the decision space to select the optimal preference elicitation strategy for the current context. However, current formulations typically treat conversational recommendation as a generic sequence modeling task, obscuring the need to explicitly learn these granular, strategic decisions. Consequently, existing approaches lack the formal structure required to drive proactive behavior systematically. To bridge this gap, we propose a new problem formulation that treats conversational recommendation as a sequential decision-making process over a structured action space, enabling the model to learn truly proactive elicitation policies.


\begin{table}[t]
\centering
\caption{Human preference on strategy-aware responses across different strategy types.}
\label{tab:strategy_better}
\begin{tabular}{lccc}
\toprule
\textbf{Strategy} & \textbf{Better (\%)} & \textbf{Worse (\%)} & \textbf{N} \\
\midrule
Item-based       & 89.34 & 10.66 & 2608 \\
Attribute-based  & 95.50 &  4.50 & 2176 \\
Hybrid           & 89.85 & 10.15 & 4198 \\
\bottomrule
\end{tabular}
\end{table}

\begin{table}[t]
  \caption{Inter-annotator agreement for strategy annotation.}
  \label{tab:consistency}
  \centering
  \begin{tabular}{lc}
    \toprule
    Metric & Agreement \\
    \midrule
    Krippendorff’s $\alpha$ (Q1) & 0.244 \\
    Krippendorff’s $\alpha$ (Q2) & 0.531 \\
    \bottomrule
  \end{tabular}
\end{table}

\section{Methodology}

\subsection{Problem Formulation}
\label{subsec:problem}

We define the conversational recommendation task as a sequential decision-making process. Let $\mathcal{U}$ and $\mathcal{V}$ denote the sets of users and items, respectively. At any given turn $t$, the dialogue context is represented as $C_t = \{u_1, s_1, \dots, u_{t-1}, s_{t-1}, u_t\}$, where $u_i$ and $s_i$ are the user and system utterances at turn $i$. 

The goal of a CRS is to learn a policy $\pi: C_t \rightarrow A_t$, where $A_t$ is a structured action representing the system's decision. To ensure explainability and control, we formulate the decision $A_t$ as a triplet $\langle a_t, \sigma_t, \phi_t \rangle$ within a multi-dimensional action space $\mathcal{A}$:
\begin{itemize}
    \item \textbf{System Action $a_t \in \mathcal{A}_{act}$}: A discrete variable indicating the high-level intent across \textit{preference elicitation}, \textit{item recommendation}, and \textit{general chat}.
    \item \textbf{Elicitation Strategy $\sigma_t \in \mathcal{A}_{pe}$}: A strategy variable indicating the decision on how to acquire user preferences (e.g., attribute-based questioning or item-based feedback).
    \item \textbf{Recommendation Objective $\phi_t \in \mathcal{A}_{rec}$}: A functional variable specifying the ranking logic or priority to select items from the item set $\mathcal{V}$ for recommendation.
\end{itemize}

With the advancements of LLMs in their superior natural language understanding, reasoning, and inference capabilities, we ground our CRS framework on an LLM backbone. This integration allows us to exploit the model's generative strengths while directing its outputs through our structured decision process. Traditional LLM-based CRSs \cite{geng2022recommendation} typically optimize a language modeling objective by maximizing the likelihood of the next token:
\begin{equation}
    \mathcal{L}_{LM} = - \sum_{i} \log P(x_i | x_{<i}, C_t)
\end{equation}
where the decision-making logic is implicitly embedded in the sequence $x$. In contrast, we propose to explicitly model the decision process by optimizing the conditional probability $P(A_t | C_t)$, where $A_t = \langle a_t, \sigma_t, \phi_t \rangle$ serves as an intermediate supervisory signal. This formulation allows the LLM to act as a central controller, integrating recommendation components within a unified end-to-end architecture:
\begin{equation}
    P(s_t | C_t) = \sum_{A_t \in \mathcal{A}} P(s_t | C_t, A_t) P(A_t | C_t)
\end{equation}
where $s_t$ is the response generated based on the explicit decision $A_t$ that indicates preferred system intents, preference elicitation strategies, and recommendations.

\begin{figure}[t] 
  \centering
  \includegraphics[
    width=0.48\textwidth,
    trim=0.5cm 17.6cm 4.5cm 0.5cm,
    clip
  ]{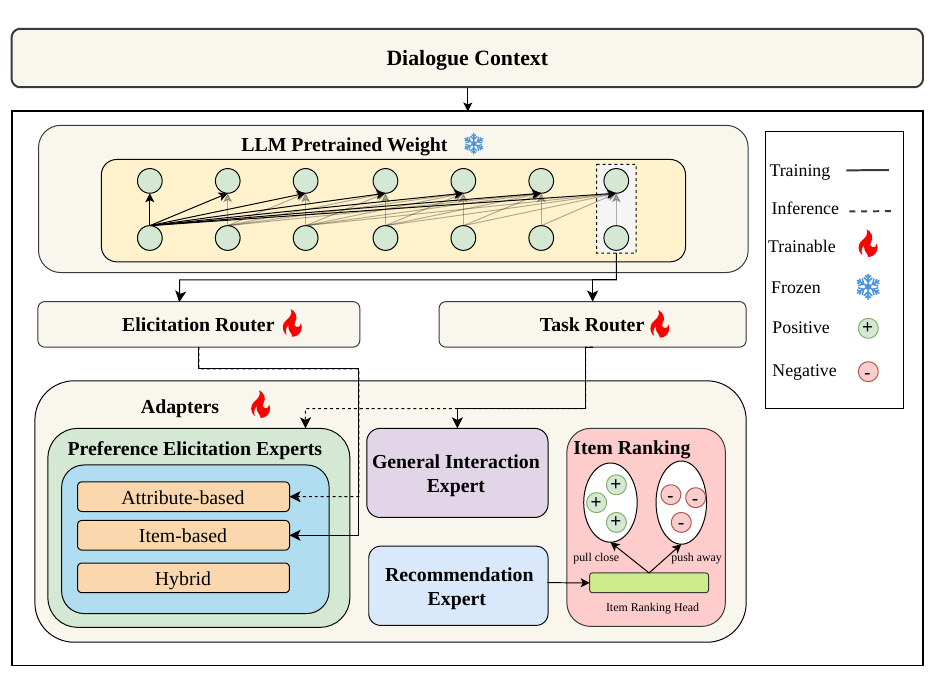}  
  \caption{Overview of our proposed conversational recommendation framework.}
  \label{fig:model}
\end{figure}

\subsection{COPE: Conversational Preference Elicitation via Mixture of Experts}
To address the complex decision modelling for conversational recommendation, we develop a novel \textbf{CO}nversational recommendation system architecture for \textbf{P}reference \textbf{E}licitation via Mixture of Experts (\textbf{COPE}). Specifically, COPE integrates multiple expert modules to explicitly model heterogeneous behaviors in CRSs. All experts share a common LLM backbone and differ only by lightweight, expert-specific parameters that are added alongside the frozen weights. Each expert is specialized for a specific task, such as preference elicitation or recommendation, and is optimized by task-specific objectives. To coordinate these experts during multi-turn interactions, we introduce a learned two-stage router (task and elicitation) that dynamically determines which expert is activated at each conversational turn, as well as the choice of strategies if preference elicitation is predicted to be necessary.
Figure~\ref{fig:model} presents an overview of the proposed architecture, illustrating how routers coordinate multiple task-specialized experts built upon a shared LLM backbone.

For the MoE component within COPE, unlike conventional Mixture-of-Experts models that primarily focus on increasing model capacity \cite{shazeer2017outrageously}, COPE is designed to explicitly separate \emph{decision execution pathways} and learn strategy-specific behavior patterns. Specifically, we designed several task-specialized experts $\Delta \Psi$, which share the same frozen LLM backbone $\Phi$. Each expert builds on the same backbone and differs only in their small set of trainable parameters. To ensure that different experts can learn their respective strategies without interfering with one another, we employ a hierarchical hard-routing scheme.
Let $C_t$ denote the dialogue context at turn $t$, and let $\mathcal{E}$ be the set of task-specialized experts. The system output $y_t$ is formulated as:
\begin{equation}
    y_t = \Phi(C_t) + \sum_{k \in \mathcal{E}} \mathbb{I}(\gamma_t = k)\cdot \Delta \Psi_k(C_t),
\end{equation}
where $\Phi(\cdot)$ represents the shared LLM backbone, $\Delta \Psi_k(\cdot)$ denotes the expert-specific adaptation for expert $k$, and $\mathbb{I}(\cdot)$ is the indicator function.
The routing variable $\gamma_t$ specifies which expert is activated at turn $t$. During training, $\gamma_t$ is directly determined by the turn-level strategy label. Only the parameters of the corresponding expert are activated and updated, while all other experts remain inactive. This teacher-forced routing stabilizes optimization and ensures that each expert focuses on learning its designated behavior without gradient interference from other CRS objectives. At inference time, $\gamma_t$ is predicted by a separately learned \textit{hierarchical router}, using dialogue context $C_t$ input, enabling dynamic expert selection without access to ground-truth labels. 

\subsubsection{Hierarchical Router}
The effectiveness of COPE depends on selecting the appropriate expert at each conversational turn. To support this, we design a hierarchical router that performs expert selection based on the dialogue context. Given the dialogue context $C_t$, we use the final hidden state of the shared frozen LLM backbone as the contextual representation:
\begin{equation}
    \mathbf{h}_{t} = \text{LLM}(C_t; \Theta_{\text{frozen}})_{[-1]}.
\end{equation}

Routing decisions are made in \textit{two stages}. At the task level, a router determines the high-level function of the current turn. In this work, we focus on preference elicitation, recommendation, and general interaction as the three major system intents for prediction. The task-level decision is computed as:
\begin{equation}
    P(a_t \mid C_t) = \text{Softmax}(W_{\text{task}} \mathbf{h}_{t} + b_{\text{task}}),
\end{equation}
where $a_t \in \{\text{Elicit}, \text{Recommend}, \text{General}\}$. Then, when preference elicitation is selected, the strategy-level router distinguishes between attribute-based, item-based, and hybrid elicitation behaviors. This design reflects our central focus on how elicitation strategies should be selected as the dialogue context evolves.
\begin{equation}
    P(\sigma_t \mid C_t, a_t=\text{Elicit}) = \text{Softmax}(W_{\text{str}} \mathbf{h}_{t} + b_{\text{str}}),
\end{equation}
where $\sigma_t \in \{\text{Attr}, \text{Item}, \text{Hybrid}\}$.

During training, expert activation is guided by ground-truth labels rather than router predictions. The router is separately trained to align routing decisions with annotated data by modelling context, without influencing which expert is activated.

\begin{algorithm}[t]
\caption{Training and Inference stage of COPE}
\label{alg:mode}
\KwIn{Dialogue turns $\{C_t\}$ with task labels $\alpha_t^*$ and strategy labels $\sigma_t^*$}
\KwOut{System response $s_t$ at each turn}

\textbf{Training Phase (Teacher-Forced Routing):} \\
\ForEach{dialogue turn $t$}{
    Extract contextual representation $\mathbf{h}_t \leftarrow \text{LLM}(C_t; \Theta_{\text{frozen}})_{[-1]}$\;
    Compute routing probabilities $P(\alpha_t \mid C_t)$ and $P(\sigma_t \mid C_t, \alpha_t)$\;
    Update router parameters using $\mathcal{L}_{\text{route}}$\;
    Determine expert index $\gamma_t$ from ground-truth labels $(\alpha_t^*, \sigma_t^*)$\;
    Activate expert $\Delta\Psi_{\gamma_t}$ and freeze all others\;
    Update expert parameters using $\mathcal{L}_{\text{sft}}$ or $\mathcal{L}_{\text{rec}}$\;
}

\vspace{0.5em}
\textbf{Inference Phase (Predicted Routing):} \\
\ForEach{dialogue turn $t$}{
    Compute $\mathbf{h}_t \leftarrow \text{LLM}(C_t; \Theta_{\text{frozen}})_{[-1]}$\;
    Predict routing decision $\hat{\gamma}_t \leftarrow \arg\max P(\alpha_t, \sigma_t \mid C_t)$\;
    Generate response $s_t$ using $\Phi(C_t) + \Delta\Psi_{\hat{\gamma}_t}(C_t)$\;
}
\end{algorithm}

\subsubsection{Multi-Task Learning}
We jointly train the router and task-specific experts using a multi-objective loss. The overall loss $\mathcal{L}_{\text{total}}$ is defined as a weighted sum of three task-specific terms:

\paragraph{Routing Loss ($\mathcal{L}_{\text{route}}$)}
We train the routers using supervised signals derived from annotated task and strategy labels.
Specifically, let $\alpha_t^*$ and $\sigma_t^*$ denote the ground-truth task label and elicitation strategy label at turn $t$, respectively.
The routing loss is defined as a weighted sum of two cross-entropy terms:
\begin{equation}
    \mathcal{L}_{\text{route}} =
    - \lambda_{\alpha} \sum_t \log P(\alpha_t^* \mid C_t)
    - \lambda_{\sigma} \sum_t \mathbb{I}(\alpha_t^* = \text{Elicit}) \log P(\sigma_t^* \mid C_t, \alpha_t^*),
\end{equation}
where the first term supervises task-level routing, and the second term is applied only when preference elicitation is selected.
The coefficients $\lambda_{\alpha}$ and $\lambda_{\sigma}$ control the relative contributions of task-level and strategy-level routing losses.

\paragraph{Expert-Specific Supervised Fine-Tuning ($\mathcal{L}_{\text{sft}}$)}
We apply supervised fine-tuning to train each task-specific expert using a standard language modeling objective.
During training, the expert activated at turn $t$ is determined by the ground-truth task or strategy label, and only the parameters of the corresponding expert are updated.

Formally, the supervised fine-tuning loss is defined as:
\begin{equation}
    \mathcal{L}_{\text{sft}} = - \sum_{i} \log P(x_i \mid x_{<i}, C_t; \Delta \Psi_{\gamma_t}),
\end{equation}
where $P(\cdot)$ is parameterized by the output distribution of the activated expert, and $\Delta \Psi_{\gamma_t}$ denotes the parameters of the expert selected at turn $t$.

\paragraph{Contrastive Recommendation Loss ($\mathcal{L}_{\text{rec}}$)}
For recommendation turns, we train the recommendation expert using a supervised contrastive ranking loss. We initialize item embeddings with text representations extracted by a pre-trained LLM, which provide semantic information about items. Given a representation produced by the activated recommendation expert from the dialogue context, the model is encouraged to get higher similarity to relevant items than to non-relevant ones. Specifically, we adopt an InfoNCE loss. Let $e_u$ denote the context representation generated by the recommendation expert, $e_{i+}$ a positive item, and $\{e_{i-}^{(j)}\}_{j=1}^{N}$ a set of negative items. The recommendation loss is defined as:
\begin{equation}
    \mathcal{L}_{\text{rec}} =
    - \log
    \frac{\exp(\text{sim}(e_u, e_{i+}) / \tau)}
    {\exp(\text{sim}(e_u, e_{i+}) / \tau)
    + \sum_{j=1}^{N} \exp(\text{sim}(e_u, e_{i-}^{(j)}) / \tau)},
\end{equation}
where $\text{sim}(\cdot)$ denotes cosine similarity and $\tau$ is a temperature hyperparameter.
The entire training process (shown in Algorithm~\ref{alg:mode}) consists of two stages: At the training stage, we train the router of COPE, perform strategy selection with ground-truth labels, and then update the marginal trainable parameters of activated experts, while at the inference stage, COPE generates responses with the activated expert selected according to the prediction of routers.

\section{Experiments}
\label{sec:experiments}

\subsection{Experimental Setup}
\label{subsec:exp_setup}

\begin{table}[t]
\centering
\caption{Key Hyperparameter Configurations.}
\label{tab:hyperparameters}
\small
\begin{tabular}{lp{4.5cm}c}
\toprule
\textbf{Notation} & \textbf{Description} & \textbf{Value} \\
\midrule
$w_{\text{sft}}$   & Supervised fine-tuning loss weight & 0.5 \\
$w_{\text{task}}$  & Action type prediction weight & 1.0 \\
$w_{\text{eli}}$   & Elicitation strategy weight & 1.0 \\
$w_{\text{rec}}$   & Recommendation ranking weight & 1.0 \\
$P_{\text{drop}}$  & LoRA dropout probability & 0.05 \\
\bottomrule
\end{tabular}
\end{table}

\paragraph{Datasets.}
We evaluate our models on the InPE dataset, which contains 6,719 turn-level samples derived from the INSPIRED dataset. Following standard practice, we split the data into training, validation, and test sets with a 6:2:2 ratio. As per the data annotation, the InPE dataset reflects a diverse distribution of dialogue acts: 46.3\% general interaction turns, 19.8\% preference elicitation turns, and 33.9\% recommendation turns. Each turn-level sample is annotated with several labels: (1) a task label indicating whether the system performs preference elicitation, recommendation, or general interaction; (2) for preference elicitation turns, a label specifying the elicitation strategy commonly preferred by annotators; (3) the target item(s) involved in recommendation turns; and (4) contrastive response pairs, one strategy-aware and one baseline—annotated to indicate which version offers a better user experience. In particular, the ``positive'' label is strictly based on human judgment. As indicated in Table~\ref{tab:strategy_better}, while the strategy-aware update is designated as the positive instance in the majority of cases, the original response is retained as the positive instance in the rare cases where the update failed to improve quality.


\textit{Evaluation Metrics.}
We conduct a comprehensive evaluation of our system from three complementary aspects: 

\noindent\textbf{$\bullet$} \textbf{(i) Recommendation Quality:} We assess recommendation performance using ranking metrics (Recall@$k$), evaluating if the system identifies the target item(s) in a top ranking. 

\noindent\textbf{$\bullet$} \textbf{(ii) Decision Quality:} We evaluate decision quality at the turn level using two accuracy metrics: \textit{Task Accuracy} for action prediction and \textit{Strategy Accuracy} for elicitation strategy selection.

\noindent\textbf{$\bullet$} \textbf{(iii) Response Quality:} We evaluate response quality using human-annotated response pairs $(y_w, y_l)$, where $y_w$ is the preferred response and $y_l$ is the less preferred one. We calculate (1) \textit{Pairwise Win Rate} by checking whether the model assigns a higher log-likelihood to the preferred response $y_w$ than to $y_l$; and (2) \textit{the Log-likelihood Margin}, measuring the difference in log-likelihood between responses $y_w$ and $y_l$:
    \begin{equation}
    \Delta_{\text{LL}}=\mathbb{E}\!\left[\frac{\log P(y_w)}{|y_w|}-\frac{\log P(y_l)}{|y_l|}\right].
    \end{equation}

\paragraph{Baselines.} To evaluate the effectiveness of our proposed framework, we compare it against a diverse set of representative conversational recommendation baselines, categorized as follows:

\noindent\textbf{$\bullet$} \textbf{Neighborhood-based methods:} We include NBCRS~\cite{xie2024neighborhood}, which performs conversational recommendation by matching test dialogues to similar training contexts and recommending items frequently associated with neighboring conversations, without relying on external knowledge bases or large models.
    
\noindent\textbf{$\bullet$} \textbf{Knowledge Graph(KG)-based methods:} We include KBRD~\cite{chen2019towards}, KGSF~\cite{zhou2020improving}, and TREA~\cite{li2023trea}, which incorporate external knowledge graphs to enhance preference modeling, dialogue understanding, and recommendation. These methods exploit structured entity relations and reasoning over dialogue history to bridge the semantic gap between natural language utterances and item-level preferences.
    
\noindent\textbf{$\bullet$} \textbf{LLM-based approaches:} We consider two categories: (i) LLM-enhanced conversational recommenders such as ReFICR~\cite{yang2024unleashing}, and (ii) Pre-trained LLMs: Qwen \cite{yang2025qwen3} and LLaMA \cite{grattafiori2024llama}, which are used as prompt-based baselines without task-specific fine-tuning.

\textit{Implementation Details.}
We implement our framework using PyTorch and the HuggingFace library. We adopt Qwen3-8B \cite{yang2025qwen3} as the backbone large language model. To maintain parameter efficiency, we employ LoRA \cite{hu2022lora} with a rank $r=8$ and $\alpha=16$. The model is fine-tuned for 1 epoch using the AdamW optimizer with a peak learning rate of $2\times10^{-5}$ and a warmup ratio of $0.03$. We use a batch size of 8 with 4 gradient accumulation steps, resulting in an effective batch size of 32. The maximum sequence length is set to 2048 tokens. All experiments are conducted on a single NVIDIA A100 (80GB) GPU using BF16 mixed-precision training. For the multi-task learning objective, the loss weights are carefully tuned as summarized in Table~\ref{tab:hyperparameters}.

\subsection{Overall Performance}
\label{subsec:strategy_results}
\begin{table}[t]
\centering
\caption{Recommendation performance on InPE Dataset.}
\label{tab:rec_results}
\setlength{\tabcolsep}{7pt}
{\small
\begin{tabular}{lccc}
\toprule
Method & R@1 $\uparrow$ & R@10 $\uparrow$ & R@50 $\uparrow$ \\
\midrule
NBCRS \cite{xie2024neighborhood} & 0.023 & 0.160 & 0.273 \\
KBRD \cite{chen2019towards} & 0.056 & 0.145 & 0.210 \\
KGSF \cite{zhou2020improving} & 0.057 & 0.165 & 0.256 \\
ReFICR \cite{yang2024unleashing} & 0.135 & 0.274 & 0.396 \\
TREA \cite{li2023trea} & 0.047 & 0.147 & 0.346 \\
Qwen3-8B \cite{yang2025qwen3} & 0.013 & 0.094 & 0.156 \\
Llama3-8B-Instruct \cite{grattafiori2024llama} & 0.031 & 0.086 & 0.148 \\
Ours  & \textbf{0.144} & \textbf{0.314} & \textbf{0.442} \\
\bottomrule
\end{tabular}
}
\end{table}

\begin{table}[t]
\centering
\caption{Pairwise evaluation on InPE Dataset.}
\label{tab:pair_results}
\setlength{\tabcolsep}{8pt}
{\small
\begin{tabular}{lcc}
\toprule
Method & Win Rate $\uparrow$ & Margin $\uparrow$ \\
\midrule
LLaMA-3.1-8B & 0.503 & -0.018 \\
Qwen3-8B & 0.452 & -0.044 \\
ReFICR & 0.551 & 0.236 \\
\midrule
\textbf{Ours} & \textbf{0.604} & \textbf{0.314} \\
\bottomrule
\end{tabular}
}
\end{table}

\begin{figure*}[t] 
  \centering
  \includegraphics[
    width=1.0\textwidth,
    trim=0.0cm 0.0cm 0.0cm 0.0cm,
    clip
  ]{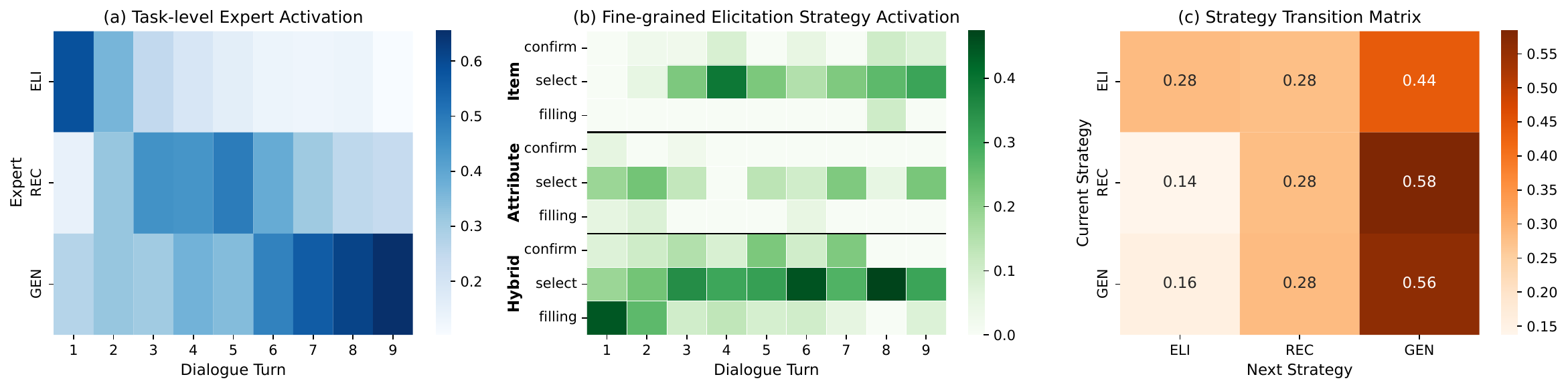}  
  \caption{Quantitative analysis of the dynamic routing mechanism. (a) Task-level expert activation probabilities across dialogue turns. (b) Fine-grained activation heatmap of ELI sub-strategies. (c) State transition matrix between different dialogue strategies (Elicitation, Recommendation, General Interaction) as determined by the router.}
  \label{fig:eli_fine_grained_activation}
\end{figure*}

Table~\ref{tab:rec_results} shows the overall recommendation performance. COPE achieves the highest recall across all $k$ values, consistently outperforming baselines from all three categories. The performance gains are particularly observable at deeper cutoffs (i.e., $k=10$ and $k=50$). 
Among all the baselines, ReFICR, which also leverages an LLM backbone, still falls short of COPE's performance. This gap highlights the critical advantage of explicitly modelling proactive preference elicitation strategies, rather than relying solely on the latent capabilities of the LLM. Furthermore, off-the-shelf prompt-based LLMs (Qwen3-8B and LLaMA3-8B-Instruct) yield substantially lower recall scores, reflecting their limited ability for recommendations without task-specific fine-tuning or guidance on proactive decision-making.  Following item-level ranking evaluation, we assessed response-level preference alignment via pairwise evaluation. Table~\ref{tab:pair_results} reports the results, grounded on the annotated comparison between strategy-aware responses and the original dialogue responses. Since response quality is evaluated based on generation likelihood, only LLM-based CRS approaches are included for comparison. We observe that prompt-based LLM baselines (LLaMA-3.1-8B and Qwen3-8B) yield win rates of around 50\% with negative margins, showing the preference estimation close to random guessing. ReFICR, the best performing baseline, achieves higher win rates with positive margins, indicating clearer preferences. However, COPE attains the highest win rate and margin among all methods, confirming the superiority of its elicitation strategy-aware approach. 

\subsection{Ablation Study}
To investigate the contribution of each specialized expert within COPE, we perform an ablation study by selectively removing the Recommendation (Rec) and Elicitation (Eli) experts. In these variants, the requests to the removed component are handled by the General expert during both training and inference, without introducing a task-specific adapter. For example, when the recommendation expert is ablated, the general expert will jointly process the system responses dealing with general chats and recommendations. Table~\ref{tab:ablation_reward} summarizes the results. The variant without any specialized experts (w/o all) achieves the highest scores on Task Accuracy, Strategy Accuracy, and Recall@10. This suggests that a single adapter is learned to support stable decision-making and retrieval signals, without the additional coordination required by multiple experts. Despite its strong performance on accuracy-based metrics, the w/o all variant performs worst in pairwise evaluation. This discrepancy indicates that while a unified model can predict labels accurately, it tends to generate generic responses that lack the nuanced, strategy-aware behaviors introduced by specialized elicitation or recommendation modules.  In contrast, removing only one expert (w/o Rec or w/o Eli) generally degrades performance.  In the full framework, experts are trained jointly and rely on complementary signals. Removing one expert disrupts this synergy, forcing the remaining specialized components to operate on unoptimized inputs from the general module. This partial setup proves less effective than either the holistic coordination of the full model or the simplicity of the fully unified variant. Ultimately, the \textbf{Full Model} provides the most robust solution, balancing competitive accuracy with superior conversational quality.

\begin{table}[!t]
\centering
\caption{Ablation study on removing specific experts.}
\label{tab:ablation_reward}
\small
\begin{tabular}{lcccc}
\toprule
\textbf{Variant} & \textbf{Task Acc} & \textbf{Eli Acc}  & \textbf{Recall@10} & \textbf{Win Rate} \\
\midrule
(1) Full      & 0.618 & 0.711 & 0.314 & 0.604 \\
(2) w/o Rec   & 0.566 & 0.623  & 0.303 & 0.598 \\
(3) w/o Eli  & 0.602 & 0.586 & 0.315 & 0.628 \\
(4) w/o all   & 0.639  & 0.756  & 0.335 & 0.525 \\
\bottomrule
\end{tabular}
\end{table}

\begin{table}[t]
\centering
\caption{Routing intervention results with different strategy executions. All settings use the same COPE backbone and differ only in how the strategy expert is selected.
}
\label{tab:routing_intervention}
\small
\begin{tabular}{lccc}
\toprule 
\textbf{Router} & \textbf{TaskAcc} & \textbf{ELI@1} & \textbf{R@10} \\
\midrule
Ground Truth        & 1.0 & 1.0 & 0.342 \\  
Prediction   & 0.617 & 0.711 & 0.314 \\
Random     & 0.333 & 0.333 & 0.051 \\
\bottomrule
\end{tabular}
\end{table}

\subsection{Mechanistic Analysis}
To better understand how COPE works, we perform a fine-grained analysis of its internal mechanisms from two aspects.

\textit{The Impact of Routing.}
Table~\ref{tab:routing_intervention} shows the results of routing interventions under different strategy execution settings. In the Ground Truth setting, the system is forced to use the correct expert (TaskAcc=1.0 and Eli@1=1.0), Recall@10 rises to 0.342. This performance gain serves as an upper bound, confirming that the specialized experts are highly effective when correctly activated.  In contrast, the standard Prediction router achieves a Task Accuracy of 0.617, leaving a clear gap relative to the Ground Truth setting. This indicates that the primary bottleneck in the current framework is the accuracy of strategy selection, rather than the intrinsic capacity of the experts themselves. The Random baseline performs poorly, with Recall@10 dropping to 0.051. This result shows that inappropriate expert selection severely degrades both recommendation and elicitation performance.

\textit{Decision Patterns across Turns.}
Following the intervention analysis above, we examine how routing decisions are distributed across dialogue turns. Figure~\ref{fig:eli_fine_grained_activation} shows the distribution of router activations across dialogue turns. Different elicitation strategies show distinct stage-wise tendencies. Attribute-based strategies are more frequently used in early turns, while item-based strategies appear more often in later turns. Hybrid strategies maintain relatively high usage throughout the dialogue, likely due to their ability to combine item- and attribute-based strategies. At the sub-strategy level, the system can ask users to respond with confirmation (confirm), choose between options (select), or add information (filling). We observe that \emph{select} actions dominate across all stages, particularly in middle and later turns, and are mainly associated with item- and hybrid-level strategies. In contrast, \emph{filling} actions occur more often in early turns and are more common at the attribute level, where they can help reduce the candidate space efficiently. \emph{Confirm} actions are less frequent overall and tend to appear in intermediate turns, primarily at the item level. These patterns may be related to differences in user cognitive load and preference coverage among sub-strategies, which may account for their different usage across dialogue stages.
Furthermore, Figure~\ref{fig:eli_fine_grained_activation}(c) shows the strategy transition matrix across dialogue turns. 
General interactions (GEN) appear as the most frequent next strategy across different current strategies.  Both recommendation (REC) and GEN show notable self-transitions, while ELI exhibits lower self-transitions and more frequent transitions to GEN.

\section{Conclusion}
This study investigates the optimal timing and selection of preference elicitation strategies to advance conversational recommendation systems. Through user studies and rigorous data annotation, we validated the limitations of static approaches and identified consistent, stage-dependent preference patterns. This effort yielded the \textbf{InPE} dataset, a resource enriched with fine-grained labels for strategy transitions. Building on these insights, we proposed \textbf{COPE}, a novel Mixture-of-Experts architecture designed to master the nuances of proactive strategy selection. Experimental results confirm that COPE significantly enhances both recommendation accuracy and strategy optimization, offering a robust framework for modeling the dynamic nature of conversational recommendation.



\bibliographystyle{ACM-Reference-Format}
\balance
\bibliography{references}

@inproceedings{priyogi2019preference,
  title={Preference elicitation strategy for conversational recommender system},
  author={Priyogi, Bilih},
  booktitle={Proceedings of the twelfth ACM International Conference on Web Search and Data Mining},
  pages={824--825},
  year={2019}
}

@inproceedings{zou2020towards,
  title={Towards question-based recommender systems},
  author={Zou, Jie and Chen, Yifan and Kanoulas, Evangelos},
  booktitle={Proceedings of the 43rd international ACM SIGIR conference on research and development in information retrieval},
  pages={881--890},
  year={2020}
}

@inproceedings{lei2020estimation,
  title={Estimation-action-reflection: Towards deep interaction between conversational and recommender systems},
  author={Lei, Wenqiang and He, Xiangnan and Miao, Yisong and Wu, Qingyun and Hong, Richang and Kan, Min-Yen and Chua, Tat-Seng},
  booktitle={Proceedings of the 13th ACM International Conference on Web Search and Data Mining},
  pages={304--312},
  year={2020}
}

@inproceedings{bledaite2015pairwise,
  title={Pairwise preferences elicitation and exploitation for conversational collaborative filtering},
  author={Bl{\'e}dait{\'e}, Laura and Ricci, Francesco},
  booktitle={Proceedings of the 26th ACM Conference on Hypertext and Social Media},
  pages={231--236},
  year={2015}
}

@inproceedings{ren2021learning,
  title={Learning to ask appropriate questions in conversational recommendation},
  author={Ren, Xuhui and Yin, Hongzhi and Chen, Tong and Wang, Hao and Huang, Zi and Zheng, Kai},
  booktitle={Proceedings of the 44th international ACM SIGIR Conference on Research and Development in Information Retrieval},
  pages={808--817},
  year={2021}
}

@inproceedings{yu2019visual,
  title={A visual dialog augmented interactive recommender system},
  author={Yu, Tong and Shen, Yilin and Jin, Hongxia},
  booktitle={Proceedings of the 25th ACM SIGKDD International Conference on Knowledge Discovery and Data Mining},
  pages={157--165},
  year={2019}
}

@inproceedings{mccarthy2010experience,
  title={Experience-based critiquing: Reusing critiquing experiences to improve conversational recommendation},
  author={McCarthy, Kevin and Salem, Yasser and Smyth, Barry},
  booktitle={International Conference on Case-Based Reasoning},
  pages={480--494},
  year={2010},
  organization={Springer}
}

@inproceedings{mccarthy2004dynamic,
  title={On the dynamic generation of compound critiques in conversational recommender systems},
  author={McCarthy, Kevin and Reilly, James and McGinty, Lorraine and Smyth, Barry},
  booktitle={International Conference on Adaptive Hypermedia and Adaptive Web-Based Systems},
  pages={176--184},
  year={2004},
  organization={Springer}
}

@article{hayati2020inspired,
  title={Inspired: Toward sociable recommendation dialog systems},
  author={Hayati, Shirley Anugrah and Kang, Dongyeop and Zhu, Qingxiaoyang and Shi, Weiyan and Yu, Zhou},
  journal={arXiv preprint arXiv:2009.14306},
  year={2020}
}

@article{gao2021advances,
  title={Advances and challenges in conversational recommender systems: A survey},
  author={Gao, Chongming and Lei, Wenqiang and He, Xiangnan and De Rijke, Maarten and Chua, Tat-Seng},
  journal={AI open},
  volume={2},
  pages={100--126},
  year={2021},
  publisher={Elsevier}
}

@inproceedings{zhou2020improving,
  title={Improving conversational recommender systems via knowledge graph based semantic fusion},
  author={Zhou, Kun and Zhao, Wayne Xin and Bian, Shuqing and Zhou, Yuanhang and Wen, Ji-Rong and Yu, Jingsong},
  booktitle={Proceedings of the 26th ACM SIGKDD international conference on knowledge discovery and data mining},
  pages={1006--1014},
  year={2020}
}

@inproceedings{christakopoulou2016towards,
  title={Towards conversational recommender systems},
  author={Christakopoulou, Konstantina and Radlinski, Filip and Hofmann, Katja},
  booktitle={Proceedings of the 22nd ACM SIGKDD international conference on knowledge discovery and data mining},
  pages={815--824},
  year={2016}
}

@article{rendle2012bpr,
  title={BPR: Bayesian personalized ranking from implicit feedback},
  author={Rendle, Steffen and Freudenthaler, Christoph and Gantner, Zeno and Schmidt-Thieme, Lars},
  journal={arXiv preprint arXiv:1205.2618},
  year={2012}
}

@inproceedings{wang2019kgat,
  title={Kgat: Knowledge graph attention network for recommendation},
  author={Wang, Xiang and He, Xiangnan and Cao, Yixin and Liu, Meng and Chua, Tat-Seng},
  booktitle={Proceedings of the 25th ACM SIGKDD International Conference on Knowledge Discovery and Data Mining},
  pages={950--958},
  year={2019}
}

@inproceedings{xie2024neighborhood,
  title={Neighborhood-based collaborative filtering for conversational recommendation},
  author={Xie, Zhouhang and Wu, Junda and Jeon, Hyunsik and He, Zhankui and Steck, Harald and Jha, Rahul and Liang, Dawen and Kallus, Nathan and McAuley, Julian},
  booktitle={Proceedings of the 18th ACM Conference on Recommender Systems},
  pages={1045--1050},
  year={2024}
}

@article{chen2019towards,
  title={Towards knowledge-based recommender dialog system},
  author={Chen, Qibin and Lin, Junyang and Zhang, Yichang and Ding, Ming and Cen, Yukuo and Yang, Hongxia and Tang, Jie},
  journal={arXiv preprint arXiv:1908.05391},
  year={2019}
}

@article{yang2025qwen3,
  title={Qwen3 technical report},
  author={Yang, An and Li, Anfeng and Yang, Baosong and Zhang, Beichen and Hui, Binyuan and Zheng, Bo and Yu, Bowen and Gao, Chang and Huang, Chengen and Lv, Chenxu and others},
  journal={arXiv preprint arXiv:2505.09388},
  year={2025}
}

@article{grattafiori2024llama,
  title={The llama 3 herd of models},
  author={Grattafiori, Aaron and Dubey, Abhimanyu and Jauhri, Abhinav and Pandey, Abhinav and Kadian, Abhishek and Al-Dahle, Ahmad and Letman, Aiesha and Mathur, Akhil and Schelten, Alan and Vaughan, Alex and others},
  journal={arXiv preprint arXiv:2407.21783},
  year={2024}
}

@article{li2023trea,
  title={Trea: Tree-structure reasoning schema for conversational recommendation},
  author={Li, Wendi and Wei, Wei and Qu, Xiaoye and Mao, Xian-Ling and Yuan, Ye and Xie, Wenfeng and Chen, Dangyang},
  journal={arXiv preprint arXiv:2307.10543},
  year={2023}
}

@inproceedings{yang2024unleashing,
  title={Unleashing the retrieval potential of large language models in conversational recommender systems},
  author={Yang, Ting and Chen, Li},
  booktitle={Proceedings of the 18th ACM Conference on Recommender Systems},
  pages={43--52},
  year={2024}
}

@inproceedings{deng2021unified,
  title={Unified conversational recommendation policy learning via graph-based reinforcement learning},
  author={Deng, Yang and Li, Yaliang and Sun, Fei and Ding, Bolin and Lam, Wai},
  booktitle={Proceedings of the 44th International ACM SIGIR Conference on Research and Development in Information Retrieval},
  pages={1431--1441},
  year={2021}
}

@inproceedings{hu2022learning,
  title={Learning to infer user implicit preference in conversational recommendation},
  author={Hu, Chenhao and Huang, Shuhua and Zhang, Yansen and Liu, Yubao},
  booktitle={Proceedings of the 45th International ACM SIGIR conference on research and development in information retrieval},
  pages={256--266},
  year={2022}
}

@book{krippendorff2018content,
  title={Content analysis: An introduction to its methodology},
  author={Krippendorff, Klaus},
  year={2018},
  publisher={Sage publications}
}

@inproceedings{he2023large,
  title={Large language models as zero-shot conversational recommenders},
  author={He, Zhankui and Xie, Zhouhang and Jha, Rahul and Steck, Harald and Liang, Dawen and Feng, Yesu and Majumder, Bodhisattwa Prasad and Kallus, Nathan and McAuley, Julian},
  booktitle={Proceedings of the 32nd ACM International Conference on Information and Knowledge Management},
  pages={720--730},
  year={2023}
}

@article{hu2022lora,
  title={Lora: Low-rank adaptation of large language models.},
  author={Hu, Edward J and Shen, Yelong and Wallis, Phillip and Allen-Zhu, Zeyuan and Li, Yuanzhi and Wang, Shean and Wang, Lu and Chen, Weizhu and others},
  journal={ICLR},
  volume={1},
  number={2},
  pages={3},
  year={2022}
}

@inproceedings{zhang2022multiple,
  title={Multiple choice questions based multi-interest policy learning for conversational recommendation},
  author={Zhang, Yiming and Wu, Lingfei and Shen, Qi and Pang, Yitong and Wei, Zhihua and Xu, Fangli and Long, Bo and Pei, Jian},
  booktitle={Proceedings of the ACM Web Conference 2022},
  pages={2153--2162},
  year={2022}
}

@misc{qwen3technicalreport,
      title={Qwen3 Technical Report}, 
      author={Qwen Team},
      year={2025},
      eprint={2505.09388},
      archivePrefix={arXiv},
      primaryClass={cs.CL},
      url={https://arxiv.org/abs/2505.09388}, 
}

@inproceedings{geng2022recommendation,
  title={Recommendation as language processing (rlp): A unified pretrain, personalized prompt and predict paradigm (p5)},
  author={Geng, Shijie and Liu, Shuchang and Fu, Zuohui and Ge, Yingqiang and Zhang, Yongfeng},
  booktitle={Proceedings of the 16th ACM Conference on Recommender Systems},
  pages={299--315},
  year={2022}
}

@article{shazeer2017outrageously,
  title={Outrageously large neural networks: The sparsely-gated mixture-of-experts layer},
  author={Shazeer, Noam and Mirhoseini, Azalia and Maziarz, Krzysztof and Davis, Andy and Le, Quoc and Hinton, Geoffrey and Dean, Jeff},
  journal={arXiv preprint arXiv:1701.06538},
  year={2017}
}

@article{hernandez2023explaining,
  title={Explaining recommendations through conversations: Dialog model and the effects of interface type and degree of interactivity},
  author={Hernandez-Bocanegra, Diana C and Ziegler, J{\"u}rgen},
  journal={ACM Transactions on Interactive Intelligent Systems},
  volume={13},
  number={2},
  pages={1--47},
  year={2023},
  publisher={ACM New York, NY}
}

@article{kostric2024generating,
  title={Generating usage-related questions for preference elicitation in conversational recommender systems},
  author={Kostric, Ivica and Balog, Krisztian and Radlinski, Filip},
  journal={ACM Transactions on Recommender Systems},
  volume={2},
  number={2},
  pages={1--24},
  year={2024},
  publisher={ACM New York, NY}
}

@inproceedings{xie2021comparison,
  title={Comparison-based conversational recommender system with relative bandit feedback},
  author={Xie, Zhihui and Yu, Tong and Zhao, Canzhe and Li, Shuai},
  booktitle={Proceedings of the 44th International ACM SIGIR Conference on Research and Development in Information Retrieval},
  pages={1400--1409},
  year={2021}
}

@inproceedings{sun2018conversational,
  title={Conversational recommender system},
  author={Sun, Yueming and Zhang, Yi},
  booktitle={The 41st international acm sigir conference on research and development in information retrieval},
  pages={235--244},
  year={2018}
}

@inproceedings{wang2022towards,
  title={Towards unified conversational recommender systems via knowledge-enhanced prompt learning},
  author={Wang, Xiaolei and Zhou, Kun and Wen, Ji-Rong and Zhao, Wayne Xin},
  booktitle={Proceedings of the 28th ACM SIGKDD conference on knowledge discovery and data mining},
  pages={1929--1937},
  year={2022}
}

@inproceedings{guo2023towards,
  title={Towards explainable conversational recommender systems},
  author={Guo, Shuyu and Zhang, Shuo and Sun, Weiwei and Ren, Pengjie and Chen, Zhumin and Ren, Zhaochun},
  booktitle={Proceedings of the 46th International ACM SIGIR conference on research and development in information retrieval},
  pages={2786--2795},
  year={2023}
}

\end{document}